\documentclass[%
 reprint,
 amsmath,amssymb,
 aps,
]{revtex4-2}

\usepackage{graphicx}% Include figure files
\usepackage{dcolumn}% Align table columns on decimal point
\usepackage{bm}% bold math
\usepackage{amsmath, amssymb}
\usepackage{hyperref}
\usepackage{float}

\usepackage[braket, qm]{qcircuit}
\usepackage{mathptmx}
\usepackage{braket}
\usepackage{amsfonts}
\usepackage{csquotes}
\usepackage{hhline}
\usepackage{relsize}
\usepackage{listings}
\usepackage{color}
\usepackage{qcircuit}
\usepackage{physics}
\usepackage{soul}
\usepackage{subcaption}
\usepackage[linesnumbered,ruled,vlined]{algorithm2e}

\usepackage{adjustbox}
\usepackage{cancel}
\begin{document}

\preprint{APS/123-QED}

\title{Recurrent Quantum Feature Maps for Reservoir Computing}% Force line breaks with \\

\author{Utkarsh Singh}
\affiliation{Department of Physics, University of Ottawa, 25 Templeton Street, Ottawa, Ontario, K1N 6N5 Canada}
\affiliation{National Research Council of Canada, 100 Sussex Drive, Ottawa, Ontario K1N 5A2, Canada}

\author{Aaron Z. Goldberg}
\affiliation{National Research Council of Canada, 100 Sussex Drive, Ottawa, Ontario K1N 5A2, Canada}

 \author{Christoph Simon}
 \affiliation{Institute for Quantum Science and Technology, Department of Physics and Astronomy, University of Calgary, Alberta T2N 1N4, Canada}
 \affiliation{Hotchkiss Brain Institute, University of Calgary, Alberta T2N 1N4, Canada}

\author{Khabat Heshami}
\affiliation{National Research Council of Canada, 100 Sussex Drive, Ottawa, Ontario K1N 5A2, Canada}
\affiliation{Department of Physics, University of Ottawa, 25 Templeton Street, Ottawa, Ontario, K1N 6N5 Canada}
 \affiliation{Institute for Quantum Science and Technology, Department of Physics and Astronomy, University of Calgary, Alberta T2N 1N4, Canada}

\begin{abstract}
Reservoir computing promises a fast method for handling large amounts of temporal data. This hinges on constructing a good reservoir--a dynamical system capable of transforming inputs into a high-dimensional representation while remembering properties of earlier data. 
In this work, we introduce a reservoir based on recurrent quantum feature maps where a fixed quantum circuit is reused to encode both current inputs and a classical feedback signal derived from previous outputs. We evaluate the model on the Mackey-Glass time-series prediction task using our recently introduced CP feature map, and find that it achieves lower mean squared error than standard classical baselines, including echo state networks and multilayer perceptrons, while maintaining compact circuit depth and qubit requirements. We further analyze memory capacity and show that the model effectively retains temporal information, consistent with its forecasting accuracy. Finally, we study the impact of realistic noise and find that performance is robust to several noise channels but remains sensitive to two-qubit gate errors, identifying a key limitation for near-term implementations.

\end{abstract}

\maketitle

%\tableofcontents

\section{Introduction \label{intro}}
Modern machine learning increasingly demands models that can process temporal data efficiently, particularly in settings where computational resources, latency, or energy consumption are constrained. While deep learning architectures have achieved remarkable success in domains such as vision, language, and control, their reliance on large-scale optimization and heavily parameterized models often limits their deployment in real-time and resource-limited environments~\cite{jaeger2004harnessing, schrauwen2007overview, tanaka2019physical}. Reservoir Computing (RC) offers a fundamentally different approach: instead of training complex recurrent dynamics, RC employs a fixed nonlinear dynamical system—the reservoir—to transform input signals into a high-dimensional representation, while only a simple linear readout is trained~\cite{maass2002real, lukovsevivcius2009reservoir}. This separation of dynamics and learning enables efficient training and has proven effective in tasks such as chaotic time-series prediction, control, and signal processing~\cite{verstraeten2007experimental, tanaka2019physical}.

Extending this paradigm into the quantum domain has led to the development of Quantum Reservoir Computing (QRC), where quantum systems serve as high-dimensional dynamical reservoirs~\cite{fujii2017harnessing}. By encoding classical inputs into quantum states and evolving them under fixed unitary dynamics, QRC leverages the exponentially large Hilbert space and intrinsic quantum correlations to enhance representational capacity. Early work demonstrated that even disordered quantum systems can match conventional recurrent neural networks on nonlinear temporal tasks~\cite{fujii2017harnessing}, motivating a wide range of implementations across physical platforms, including nuclear spin ensembles~\cite{negoro2018machine}, continuous-variable systems~\cite{nokkala2021gaussian}, superconducting qubits~\cite{chen2020temporal, yasuda2023quantum}, and large-scale neutral-atom processors~\cite{quera2023scaling}. More recent studies have further explored engineered dissipation and analog quantum dynamics to improve memory and scalability~\cite{sannia2024dissipation, govia2021quantum}. These developments position QRC as a promising framework for temporal learning on near-term quantum hardware.

Even with these advances, QRC still struggles with memory retention after measurement, as quantum observations collapse the state and break the temporal links needed for sequence processing~\cite{mujal2023time, murauer2025feedback}. Workarounds like reinitialization~\cite{kobayashi2024feedback}, mid-circuit resets, feedback-based schemes~\cite{PRXQuantum_feedback, gonon2026feedbackdrivenrecurrentquantumneural}, or weak measurements~\cite{monomi2025weak} help, but often increase complexity. Recent studies instead embrace dissipation and noise as useful features—amplitude damping and loss have been shown to enhance memory and task performance~\cite{sannia2024dissipation, domingo2023noise}. Other strategies restrict memory artificially to balance efficiency and relevance~\cite{cindrak2024memory}. Together, these developments aim to preserve QRC’s expressivity while making it practical for real-time, hardware-compatible implementations.

In parallel, quantum feature maps and kernel methods~\cite{goto2021universal, matsumoto2025iterative}—originally developed for static learning tasks—have been shown to possess high expressibility and universal approximation capabilities. These circuits encode classical data into quantum states via fixed, parameterized unitaries and have proven effective in kernel-based quantum classifiers. However, their potential as dynamical systems remains underexplored. Given their ability to map data into structured, high-dimensional quantum feature spaces, an open question is whether such circuits can be repurposed as reservoirs for temporal information processing—especially if coupled with mechanisms that introduce memory and recurrence.

In this work, we address this question by proposing a feedback-driven quantum reservoir architecture based on a reusable quantum feature map circuit. The core idea is to repurpose the fixed quantum circuit by encoding both current inputs and a feedback signal—derived from previous outputs—into separate parts of the circuit. The circuit is divided into two halves: the first encodes a sliding window of input data \(\{\mathbf{x}_t, \dots, \mathbf{x}_{t+\tau}\}\), while the second half encodes a classically computed feedback term from the prior output, scaled by a tunable strength \(\alpha\). This design introduces temporal recurrence into the system without requiring any mid-circuit measurements or resets, thereby preserving the fading-memory property and ensuring linear runtime \(\mathcal{O}(L)\). By integrating structured feedback into a fixed quantum circuit, our method bridges the gap between quantum kernel methods and reservoir computing, yielding a compact and expressive temporal learning model.

We evaluate our quantum reservoir architecture on the Mackey-Glass chaotic time-series prediction task. Across a range of delay parameters~\(\tau\), our model consistently  performs on par or better than classical reservoir computers, multilayer perceptrons (MLPs), and linear regression baselines in terms of mean squared error (MSE). We systematically explore how feedback strength~\(\alpha\), entanglement, and circuit parameters affect the model’s memory capacity and predictive performance. Dynamical stability is confirmed via the echo state property (ESP), and fading memory behavior is validated through standard memory capacity tests. 

In summary, this work introduces a versatile quantum reservoir model that simultaneously achieves recurrence, expressivity, and interpretability. By unifying kernel-based quantum learning with dynamic feedback architectures, we take a step toward general-purpose, compact quantum models for real-time, temporal processing. The rest of this paper is organized as follows. In Sec.~\ref{sec:rc}, we present a
detailed background on reservoir computing, including the echo state property and memory capacity. Sec.~\ref{sec:qrc_} introduces the proposed feedback-driven quantum reservoir architecture, along with the datasets and feature maps used in this work. Sec.~\ref{sec:results} presents the experimental results, including performance evaluation, noise analysis, and the role of feedback and entanglement in the reservoir dynamics.

\section{Reservoir Computing \label{sec:rc}}
Reservoir computing leverages a fixed, randomly initialized dynamical system—referred to as the \emph{reservoir}—to nonlinearly embed input sequences into a high-dimensional state space, where temporal dependencies can be extracted via simple linear readout mechanisms~\cite{jaeger2001echo, maass2002real, lukosevicius2009reservoir}. A conceptual illustration of this framework is shown in Fig.~\ref{fig1}.

\begin{figure}
    \centering
    \includegraphics[width=\linewidth]{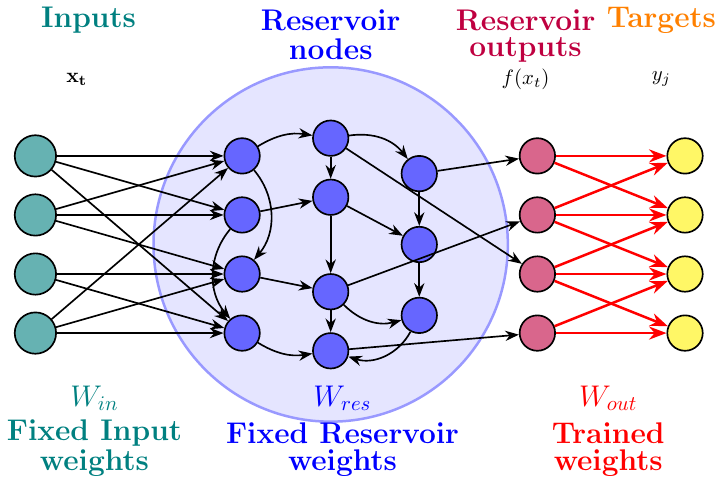}
    \caption{
        Schematic representation of the classical reservoir computing framework. 
        The input vector $\mathbf{x}_t$ is projected into a high-dimensional dynamical space by a recurrent network of fixed, randomly connected internal nodes (the reservoir), characterized by weights $\mathbf{W}_{\mathrm{in}}$ and $\mathbf{W}_{\mathrm{res}}$. 
        The resulting reservoir states are then linearly mapped to the target outputs $\mathbf{y}_j$ via a trainable readout layer with weights $\mathbf{W}_{\mathrm{out}}$. 
        Only the output layer is optimized during training, while the reservoir dynamics remain untrained, enabling efficient learning of complex temporal patterns.
    }
    \label{fig1}
\end{figure}

The mathematical formulation of an echo state network (ESN), the most widely studied reservoir computing architecture, can be expressed as follows. Let the input at time step $t$ be denoted by $\mathbf{x}_t \in \mathbb{R}^K$, where $K$ is the input dimension. The internal reservoir state vector $\mathbf{h}_t \in \mathbb{R}^N$ (representing the activations of the $N$ reservoir nodes) evolves according to~\cite{jaeger2001echo}:
\begin{equation}
\mathbf{h}_{t+1} = f\left(\mathbf{W}_{\mathrm{res}} \mathbf{h}_t + \mathbf{W}_{\mathrm{in}} \mathbf{x}_{t+1} + \mathbf{W}_{\mathrm{fb}} \mathbf{y}_t\right),
\label{eq:esn_update}
\end{equation}
where $\mathbf{W}_{\mathrm{res}} \in \mathbb{R}^{N \times N}$ defines the recurrent connections within the reservoir, $\mathbf{W}_{\mathrm{in}} \in \mathbb{R}^{N \times K}$ encodes the fixed input coupling, and $\mathbf{W}_{\mathrm{fb}} \in \mathbb{R}^{N \times L}$ governs optional feedback from the output $\mathbf{y}_t \in \mathbb{R}^L$. The function $f(\cdot)$ is typically a nonlinear activation such as the hyperbolic tangent~\cite{jaeger2004harnessing}.

The final output is computed by a linear readout function that maps the current reservoir state (and optionally the input) to the prediction:
\begin{equation}
\mathbf{y}_{t+1} = \mathbf{W}_{\mathrm{out}} \begin{bmatrix} \mathbf{h}_{t+1} \\ \mathbf{x}_{t+1} \end{bmatrix},
\label{eq:esn_output}
\end{equation}
where $\mathbf{W}_{\mathrm{out}} \in \mathbb{R}^{L \times (N + K)}$ is learned via standard regression techniques. Crucially, only $\mathbf{W}_{\mathrm{out}}$ is optimized during training, making the approach computationally efficient and well-suited for time-series tasks.

\subsection{Echo State Property and Fading Memory}

For a reservoir to be useful, its dynamics must be both stable and input-driven. Two closely related concepts capture this requirement: the echo state property (ESP) and the fading memory property. The ESP states that, for a given input sequence, the reservoir state should eventually be uniquely determined by the input history rather than by the initial condition~\cite{jaeger2001echo,yildiz2012re}. In other words, two trajectories driven by the same input should converge after a transient, even if they start from different initial states. The fading memory property complements this by requiring that the influence of past inputs decays with time, so that recent inputs affect the current state more strongly than distant ones~\cite{boyd1985fading,grigoryeva2018echo,goncalves2020reservoir}. Together, these properties ensure that the reservoir acts as a stable causal filter with finite effective memory.

In this work, we use the ESP and fading memory in their operational sense: the reservoir should forget its initialization, remain driven by the input stream and feedback signal, and retain only a finite but useful memory of the past.

\subsection{Memory Capacity and Computational Power}

The memory capacity of a reservoir computing system quantifies its ability to reconstruct past inputs from current reservoir states. Jaeger introduced the linear memory capacity as a measure of how well an ESN can linearly reconstruct delayed versions of its input~\cite{jaeger2001short}. For a scalar input $u(t)$, the linear memory capacity is defined as:

\begin{equation}
MC = \sum_{k=1}^{\infty} MC_k
\label{eq:memory_capacity}
\end{equation}
where $MC_k$ is the capacity to reconstruct the input delayed by $k$ time steps:
\begin{equation}
    MC_k = \frac{\left(\mathrm{cov}(x_{t-k}, \hat{x}_k(t))\right)^2}
{\mathrm{var}(x_{t-k}) \, \mathrm{var}(\hat{x}_k(t))}
\end{equation}
and $\hat{x}_k(t)$ is the best linear reconstruction of $x_{t-k}$ from the reservoir state $\mathbf{h}_t$~\cite{jaeger2001short}.

\section{Quantum Reservoir Architecture \label{sec:qrc_}}

We propose a quantum reservoir based on a static quantum feature map circuit, commonly used in kernel-based quantum machine learning. This reservoir consists of a two-part quantum circuit: the \textbf{left half} applies a parameterized feature map \( U(\mathbf{x}_t) \), which encodes a sliding window of time-series data \( \mathbf{x}_t = [x_{t-\tau}, \dots, x_t]\), while the \textbf{right half} applies the inverse circuit \( U^\dagger(\mathbf{z}_{t-1}) \), encoding a feedback vector derived from the output of the previous time step.

\begin{figure*}[t]
    \centering
    \includegraphics[width=0.8\textwidth]{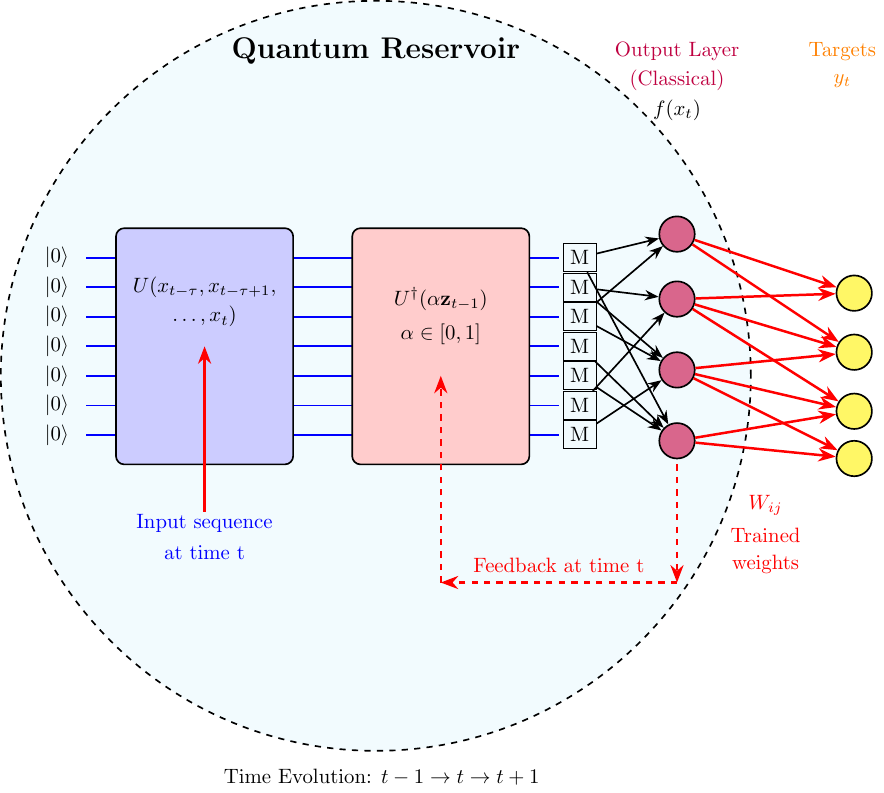}
    \caption{Schematic of proposed feedback-driven quantum reservoir computing model. At each timestep \( t \), a windowed input sequence \( [x_{t-\tau}, \dots, x_t] \) is encoded into the left half of a quantum feature map circuit \( U(\cdot) \), while the right half applies the inverted circuit \( U^\dagger(\cdot) \) using a feedback vector derived from the previous reservoir output. The feedback is modulated by a scaling parameter \( \alpha \in [0, 1] \), with \( \alpha = 1 \) corresponding to full feedback and \( \alpha = 0 \) to input-only evolution. The quantum circuit is initialized in \(|0\rangle^{\otimes n}\), and measurement outcomes are collected to produce a classical output vector, which goes to the regression model to generate the prediction \( y_t \).
}
    \label{fig1}
\end{figure*}

The feedback signal is constructed from single-qubit expectation values obtained from the circuit output distribution. Let $p_t(s)$ denote the probability of observing bitstring $s \in \{0,1\}^{n_q}$ at time step $t$, where $n_q$ is the number of qubits. The $i$-th component of the feedback vector is defined as
\begin{equation}
    z_{t,i} = \sum_{s \in \{0,1\}^{n_q}} p_t(s)\, (-1)^{s_i},
\label{eq:feedback_expectation}
\end{equation}
where $s_i$ denotes the $i$-th bit of $s$. This is equivalent to the expectation value $z_{t,i} = \langle Z_i \rangle_t$ of the Pauli-$Z$ operator on qubit $i$. Collecting all qubit expectations gives the feedback vector
\begin{equation}
    \mathbf{z}_t = \bigl(z_{t,1}, z_{t,2}, \dots, z_{t,n_q}\bigr)^\top \in \mathbb{R}^{n_q}.
\end{equation}
The feedback is then scaled by a feedback strength parameter $\alpha \in [0,1]$:
\begin{equation}
   \tilde{\mathbf{z}}_t = \alpha \mathbf{z}_t.
    \label{eq:scaled_feedback} 
\end{equation}

The resulting reservoir can be viewed as a discrete-time, input-driven quantum dynamical system. At each time step, the circuit applies a composite unitary

\begin{equation}
U_t = U^\dagger\!\bigl(\Pi(\tilde{\mathbf{z}}_{t-1})\bigr)\, U(\mathbf{x}_t),
\end{equation}
where $\Pi(\cdot)$ denotes a padding operation that matches the dimensionality of the feature map parameters. The quantum state is initialized as
\begin{equation}
\rho_0 = |0\rangle\langle 0|^{\otimes n_q},
\end{equation}
and evolves as
\begin{equation}
\rho_t = U_t \rho_0 U_t^\dagger.
\end{equation}

Measurement in the computational basis produces a probability distribution
\begin{equation}
p_t(s) = \langle s | \rho_t | s \rangle, \quad s \in \{0,1\}^{n_q}.
\end{equation}

Two different representations are then extracted from this distribution. The feedback signal is given by the expectation values in Eq.~\ref{eq:feedback_expectation}, while the regression features are constructed by selecting a subset of the measurement probabilities:
\begin{equation}
\mathbf{r}_t = \bigl(p_t(s_1), \dots, p_t(s_{\lfloor \lambda 2^{n_q} \rfloor})\bigr).
\end{equation}

Here, \( \lambda \in (0, 1] \) determines the proportion of the quantum output used for training. Specifically, if the full output vector has dimension \(m\), only the first \( \lfloor \lambda m \rfloor \) components are used as input to the regression model. This allows us to adjust the dimensionality of the learning model without modifying the quantum circuit depth or qubit count.

The final prediction is obtained via a linear readout
\begin{equation}
\hat{y}_t = \mathbf{w}_{\mathrm{out}}^\top \mathbf{r}_t + b,
\end{equation}
where the target is defined as $y_t = x_{t+h}$.

For CPMap, the number of parameters required by the second half of the circuit (feedback section) exceeds the number of qubits. In this case, the feedback vector \( \tilde{\mathbf{z}}_t \) is padded with zeros via $\Pi(\cdot)$ to match the required input dimension. In contrast, for the ZZFeatureMap, no padding is required, as the number of parameters matches the number of qubits.

\renewcommand{\thealgocf}{} 
\begin{algorithm}[htbp]
\caption{Feedback-Driven Quantum Reservoir Computing}
\label{alg:qrc_feedback}
\DontPrintSemicolon
\KwIn{Time-series input $\{x_t\}_{t=1}^{T}$, feature map $U(\cdot)$, window size $\tau$, feedback strength $\alpha$, output fraction $\lambda$, prediction horizon $H$}
\KwOut{Predicted outputs $\{\hat{y}_t\}_{t=\tau+1}^{T-H}$}

Initialize feedback vector $\tilde{\mathbf{z}}_0 \leftarrow \mathbf{0} \in \mathbb{R}^{n_q}$\;

\For{$t = \tau+1$ \KwTo $T - H$}{
    Create input window: $\mathbf{x}_t = [x_{t-\tau}, \dots, x_t]$\;

    Encode input with feature map: apply $U(\mathbf{x}_t)$ to left half of the circuit\;

    Encode feedback from previous step: apply $U^\dagger(\tilde{\mathbf{z}}_{t-1})$ to right half\;

    Execute full quantum circuit and measure output distribution $\mathbf{p}_t$\;

    Compute feedback vector $\mathbf{z}_t$ using Eq.~\ref{eq:feedback_expectation}\;

    Update feedback: $\tilde{\mathbf{z}}_t = \alpha \cdot \mathbf{z}_t$\;

    % Select first $\lfloor \lambda \cdot |\mathbf{p}_t| \rfloor$ amplitudes from $\mathbf{p}_t$ (sorted lexicographically)\;

    Construct regression features:
    $\mathbf{r}_t = (p_t(s_1), \dots, p_t(s_{\lfloor \lambda 2^{n_q} \rfloor}))$;

    Store selected components as quantum reservoir state vector for training\;

    Store $(\mathbf{r}_t, y_t)$ where $y_t = x_{t+H}$\;
    
}

Train regression model: $\hat{y}_t = \mathbf{w}_{\mathrm{out}}^\top \mathbf{r}_t + b$\;

\Return predicted sequence $\{\hat{y}_t\}_{t=\tau+1}^{T-H}$
\end{algorithm}

We also consider an alternative feedback mechanism, referred to as \emph{full-state feedback}. In this setting, instead of computing single-qubit expectation values, the feedback signal is constructed directly from the measurement probability distribution. Specifically, we take the first $n$ components of the probability vector in lexicographic order, where $n$ matches the input dimension, and scale them by the feedback strength parameter $\alpha$. This vector is then used as the feedback input to the circuit.

This approach avoids explicit computation of single-qubit expectation values and reduces classical post-processing overhead. Details and results are provided in the Supplementary Material.

One must note that all the results presented in the main text of this work are obtained using the feedback mechanism described in Algorithm~\ref {alg:qrc_feedback}. 

\subsection{Benchmark Dataset: Mackey–Glass System}

To evaluate the temporal forecasting performance of our quantum reservoir model, we employ the well-known Mackey–Glass chaotic time series~\cite{mackey1977oscillation}. Originally introduced to model physiological blood flow dynamics, the Mackey–Glass system is governed by the following nonlinear delay differential equation:
\begin{equation}
\frac{dx(t)}{dt} = b \frac{x(t - \tau)}{1 + x(t - \tau)^n} - c x(t).
\label{eq:mackeyglass}
\end{equation}
where \( b, c, \tau, n \) are positive constants. For appropriate parameter values (e.g., \( b = 0.2, c = 0.1, n = 10, \tau = 17 \)), this system exhibits deterministic chaos, characterized by sensitivity to initial conditions and complex temporal correlations. These properties make it a canonical benchmark for evaluating memory and prediction capacity in recurrent and reservoir computing models.

To simulate the system, we numerically integrate Eq.~\eqref{eq:mackeyglass} using a Runge–Kutta method with a discretization step \( \Delta t \), and sample points at uniform intervals to form a univariate time series \( \{x_t\} \). The series is normalized to the interval \([0, 1]\) or \([0, \pi]\), depending on the input encoding scheme of the quantum feature map.

\subsection{Feature Maps}

Our quantum reservoir is constructed using fixed, parameterized circuits originally designed for quantum kernel methods. Specifically, we incorporate two types of feature maps: the CPMap~\cite{Singh2025} and the standard ZZFeatureMap~\cite{havlivcek2019supervised}.

\paragraph{CPMap:} 
The CPMap is a structured, resource-efficient quantum feature map inspired by the architectural layout of quantum convolutional neural networks (QCNNs). It enables the encoding of a high number of classical features onto a limited number of qubits by using unitary transformations that coherently compress information. The CPMap alternates between data-encoding rotation layers and carefully designed two-qubit entangling blocks that focus features from multiple qubits into fewer ones. As a result, it supports hierarchical feature encoding and can embed approximately \( 2n \) classical features using only \( n \) qubits. More importantly, CPMap significantly reduces quantum resource requirements compared to standard maps, requiring only \( \sim 9F/2 \) CNOT gates to encode \( F \) features—far fewer than the quadratic CNOT scaling of the ZZFeatureMap. In this study, CPMap serves as the default circuit for both the input and feedback encoding in our quantum reservoir.

\paragraph{ZZFeatureMap:} For comparison, we also evaluate our architecture using the standard ZZFeatureMap~\cite{havlivcek2019supervised}, which encodes inputs via single-qubit \( R_Z \) rotations followed by pairwise controlled-Z entangling gates. The ZZFeatureMap is known for its simplicity, symmetry, and suitability for quantum kernel estimation. Its inclusion allows us to benchmark the performance of CPMap against an established alternative.

Each of these circuits is used in the feedback-reservoir architecture, where the left half of the circuit encodes the input sequence and the right half encodes the feedback signal via a daggered copy of the same circuit.

\section{Results and Analysis \label{sec:results}}
\subsection{Performance on Mackey-Glass Dataset}
We evaluate the forecasting performance of our feedback-based quantum reservoir computing (QRC) architecture on the Mackey--Glass time series with a delay parameter of \(\tau = 17\), which is widely recognized in the literature for generating rich chaotic dynamics. The prediction task is configured with a window size of 20 and a prediction horizon of 20, forming a moderately long-term forecasting problem. Input sequences are normalized using a standard scaling procedure, and the dataset is partitioned into 7,500 training and 2,500 test samples. For this experiment, we focus on the CPMap-based quantum circuit, as simulating the ZZFeatureMap for this setup becomes computationally prohibitive given the circuit depth and dataset size.

\begin{figure}[!t]
    \centering
    \includegraphics[width=\linewidth]{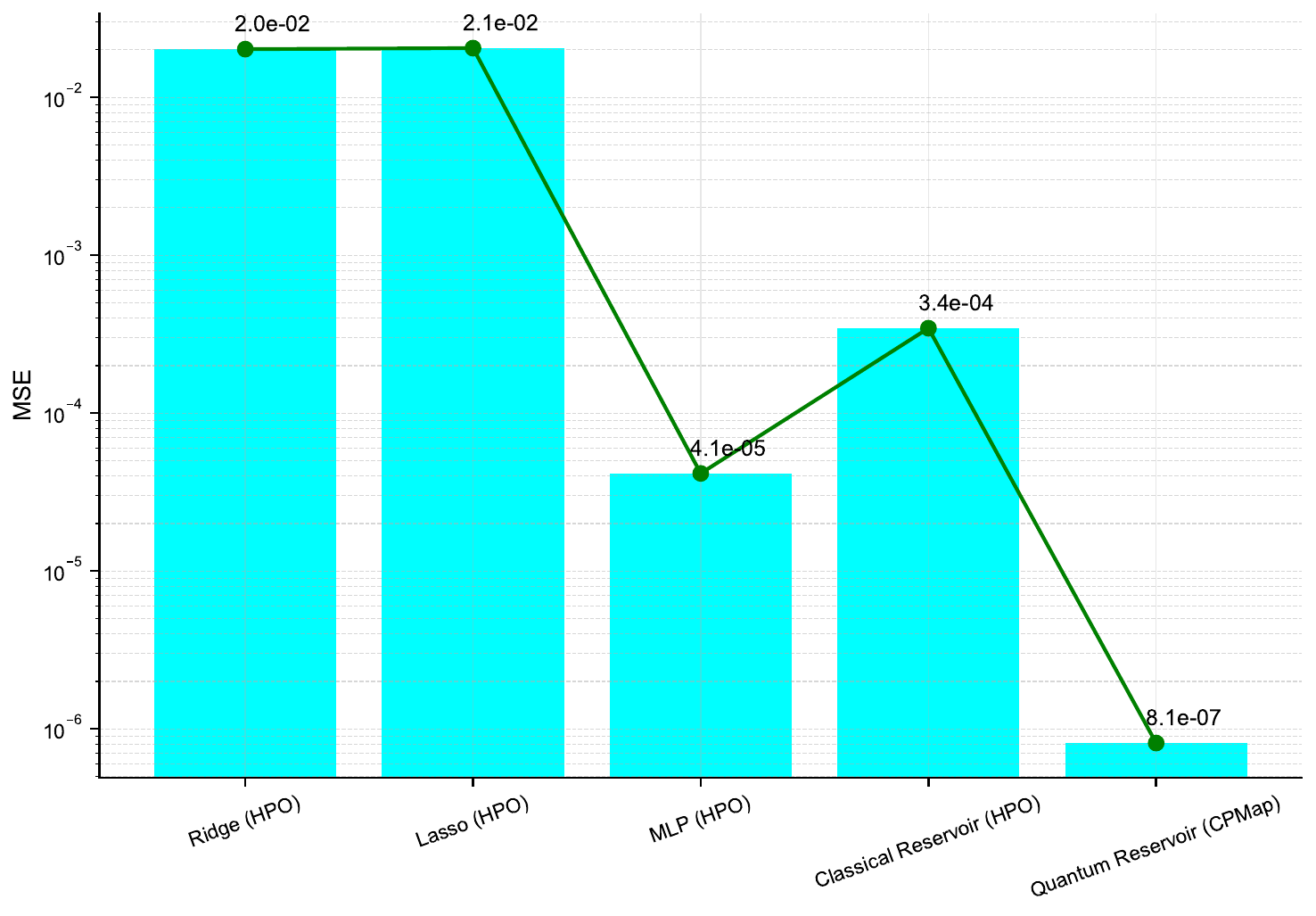}
    \caption{Model comparison on the Mackey--Glass dataset at \(\tau = 17\), window size = 20, prediction horizon = 20. Quantum reservoir achieves the lowest MSE despite no hyperparameter tuning.}
    \label{fig:model_comparison}
\end{figure}

Despite using a fixed quantum circuit with no internal trainable parameters and no hyperparameter optimization, the QRC model outperforms a range of classical baselines. As shown in Figure~\ref{fig:model_comparison}, it achieves the lowest mean squared error (MSE) among all tested models, including ridge regression, lasso regression, multilayer perceptrons (MLPs), and a hyperparameter-tuned echo state network (ESN). The ESN’s parameters—including spectral radius, input scaling, and regularization—were optimized via grid search, whereas the QRC model remains untouched after initialization. This result highlights the expressive power of the CPMap circuit when integrated into a feedback loop, enabling efficient temporal encoding and memory retention through purely unitary evolution.

\begin{figure}[!t]
    \centering
    \includegraphics[width=\linewidth]{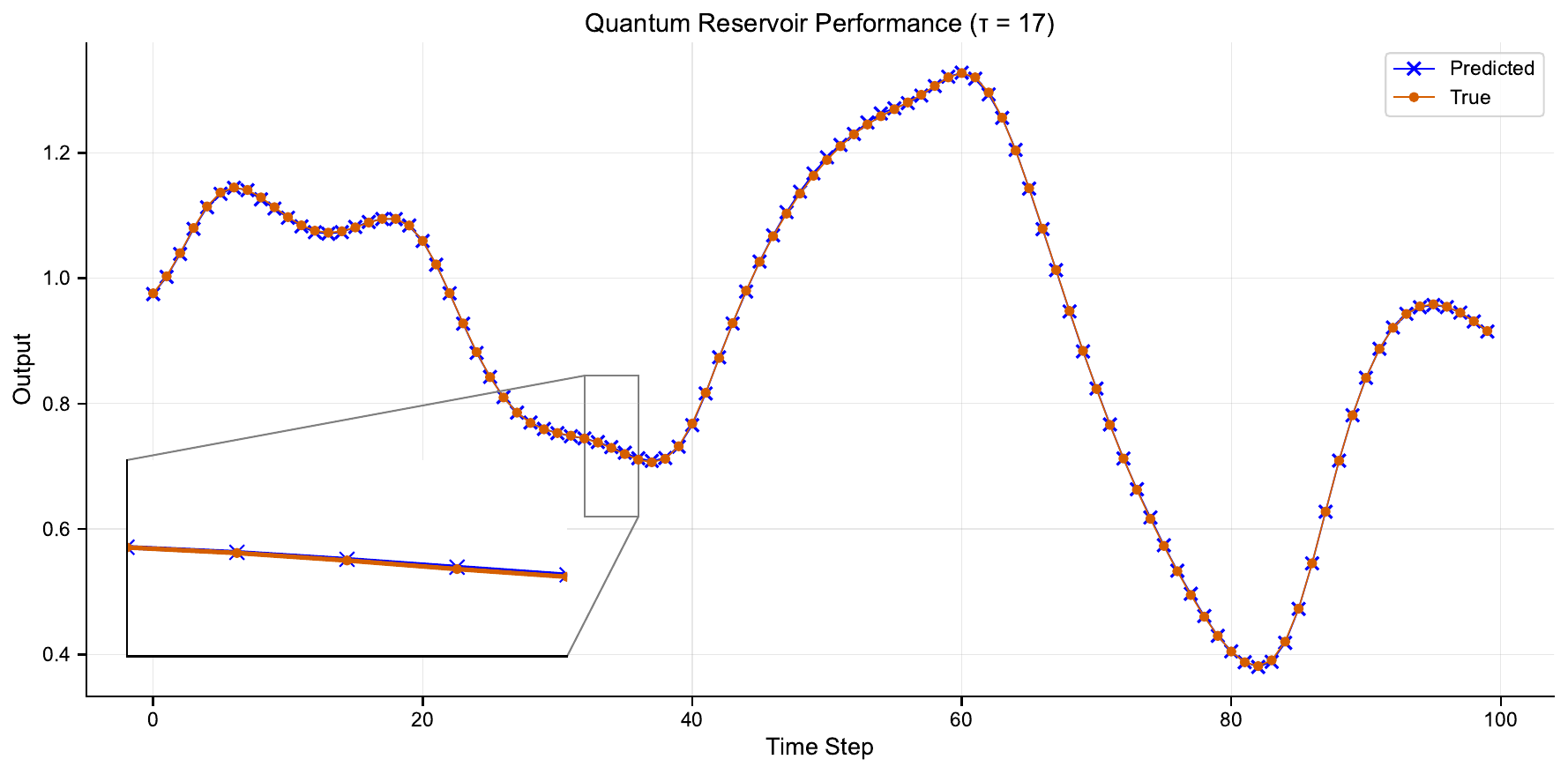}
    \caption{Predicted vs. true signal for the quantum reservoir over 100 test time steps. The reservoir captures the chaotic dynamics with high fidelity and stability.}
    \label{fig:mg_pred_trace_tau17}
\end{figure}

Figure~\ref{fig:mg_pred_trace_tau17} illustrates the qualitative forecasting behavior of the quantum reservoir over a 100-step interval in the test set. The model captures both the fast oscillations and slower amplitude modulations of the target signal, closely tracking the chaotic dynamics without significant drift or cumulative error. These results demonstrate the reservoir’s capacity to produce stable, accurate forecasts across extended time intervals.

\begin{figure}[!t]
    \centering
    % Top Figure
    \includegraphics[width=\linewidth]{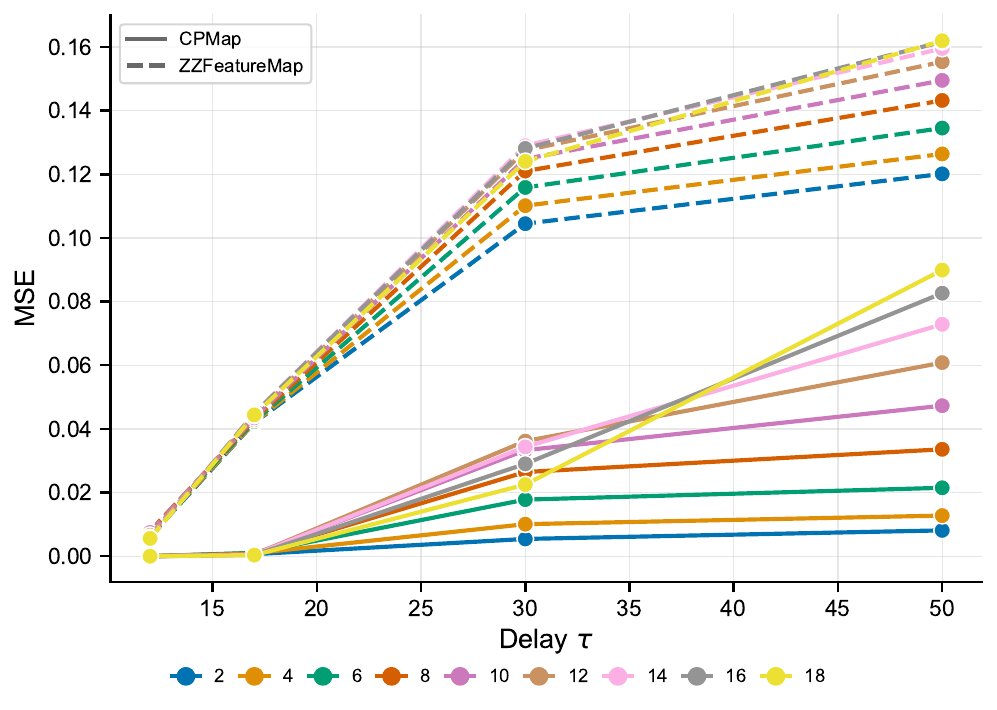}
    \\[1ex] % Adds a vertical gap between images
    
    \caption{Mean squared error (MSE) across Mackey--Glass delays \(\tau = 15\) to \(50\), for various prediction horizons, and two different feature maps. The CPMap consistently outperforms the ZZFeatureMap across delays and horizons.}
    \label{fig:mse_vs_tau_fmaps}
\end{figure}

To further evaluate the robustness of our quantum reservoir model under varying temporal dependencies, we examine performance across a range of Mackey--Glass delay parameters \(\tau \in \{12, 17, 30, 50\}\) for different prediction horizons. Figure~\ref{fig:mse_vs_tau_fmaps} shows the mean squared error (MSE) for each combination of delay and prediction horizon, separately for CPMap and ZZFeatureMap. As expected, the forecasting error generally increases with both \(\tau\) and the horizon, reflecting the increasing memory demands of the task. However, the CPMap consistently yields lower MSE across most settings, demonstrating superior robustness in retaining relevant temporal information.

Notably, while both feature maps experience degradation in accuracy at high delays and long horizons, the ZZFeatureMap exhibits more pronounced performance loss, particularly for horizons greater than 10. This contrast highlights the advantage of CPMap in encoding both current input and feedback in a more expressive and noise-resilient manner. These trends remain stable across multiple experimental runs, reinforcing the generalizability of the observed advantage.

We also analyze the influence of feedback strength ($\alpha$) and readout truncation ($\gamma$), showing that optimal performance arises from a balance between recurrence and the readout information. Supporting results and figures are provided in the Supplementary Material. 

To further understand the performance, we examine the dynamical properties of the quantum reservoir. We find that the system exhibits a finite but effective memory capacity, retaining information over a limited temporal horizon before saturating. At the same time, the reservoir satisfies the echo state property (ESP), as trajectories initialized from different states converge under identical inputs after a short time. A detailed quantitative analysis of memory capacity and ESP verification is provided in the supplementary material.

\subsection{Entanglement, Memory Capacity, and Prediction Error}

To investigate the sensitivity of the quantum reservoir’s dynamics to circuit-level hyperparameters, we performed a controlled sweep over a single circuit parameter, denoted $\theta_i$, while keeping the other five parameters fixed. For each value of $\theta_i$ sampled uniformly from the interval $[-\pi/2, \pi/2]$, we evaluated three key quantities: the short-term memory capacity (green, right axis), the reservoir’s predictive performance as measured by mean square error (MSE; red, right axis), and the average single-qubit entanglement entropy (dashed red, left axis), computed by tracing out each qubit and averaging the resulting von Neumann entropies.

\begin{figure}[!t]
    \centering
    \includegraphics[width=\linewidth]{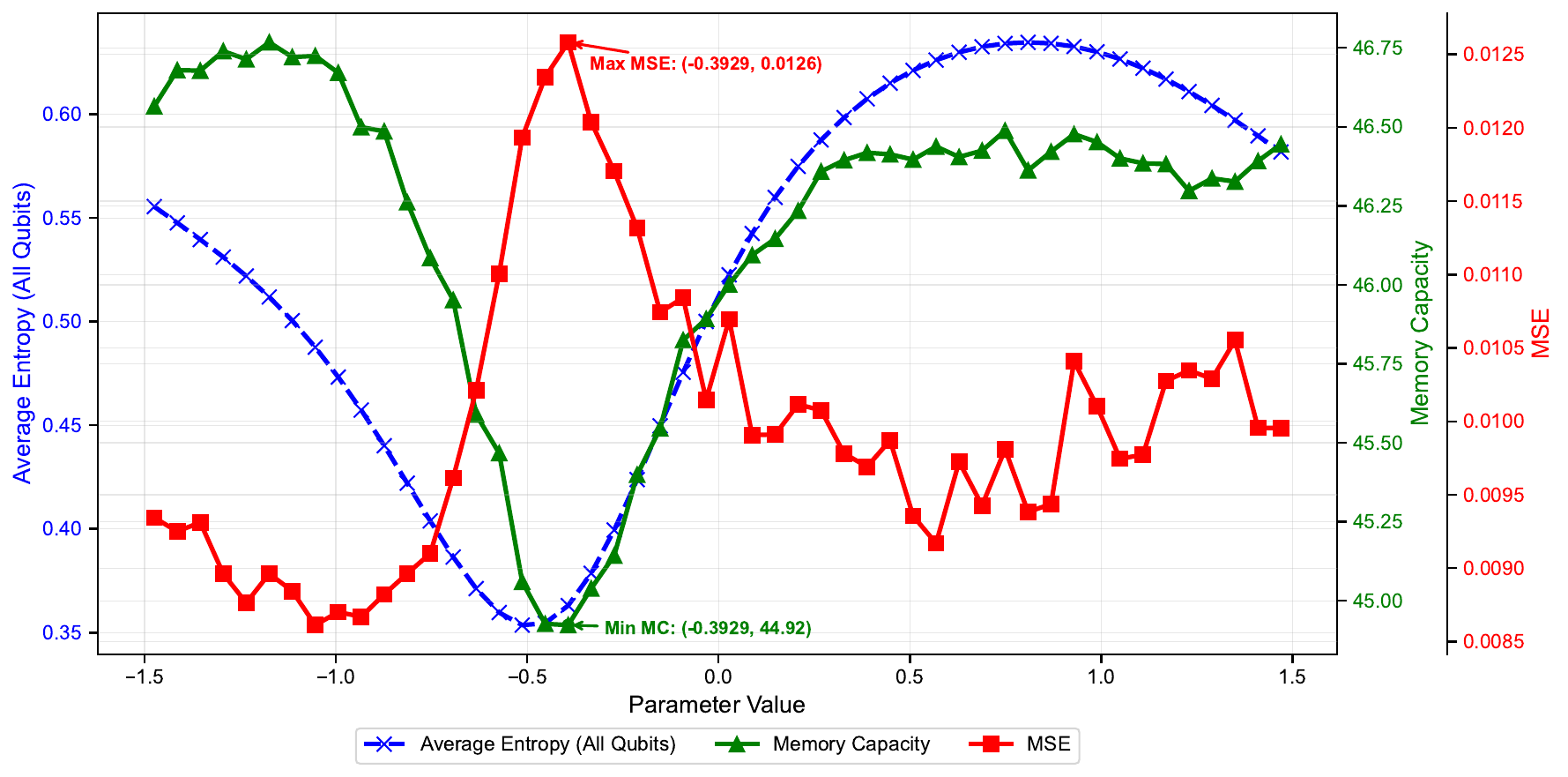}
    \caption{
    Impact of quantum circuit parameter $\theta_i$ on the reservoir’s entanglement structure, memory capacity, and predictive performance. A single parameter $\theta_i \in [-\pi/2, \pi/2]$ is swept while keeping all other circuit parameters fixed. The average single-qubit entanglement entropy (dashed blue) is computed across all qubits, the memory capacity (green) is derived from the explicit STM formulation, and MSE (solid red) quantifies forecasting error. The results reveal a structured dependence of the reservoir’s computational behavior on $\theta_i$, with distinct parameter regions exhibiting enhanced memory retention, moderate entanglement, and low prediction error.
    }
    \label{fig:entropy_memory_rmse}
\end{figure}

As shown in Fig.~\ref{fig:entropy_memory_rmse}, these three metrics exhibit nontrivial and correlated dependence on the quantum circuit configuration. Notably, regions of high memory capacity tend to coincide with improved prediction accuracy, consistent with theoretical expectations from reservoir computing. The entanglement entropy profile reveals a complementary structure: excessively high or low entanglement appears to degrade performance, suggesting an optimal intermediate regime where the circuit remains expressive yet stable. These findings highlight the sensitivity of quantum reservoir dynamics to circuit-level parameters and underscore the utility of entanglement entropy as a diagnostic for task-relevant quantum behaviour.

\subsection{Result of noisy simulation:}

\begin{figure}[!t]
    \centering
    \includegraphics[width=\linewidth]{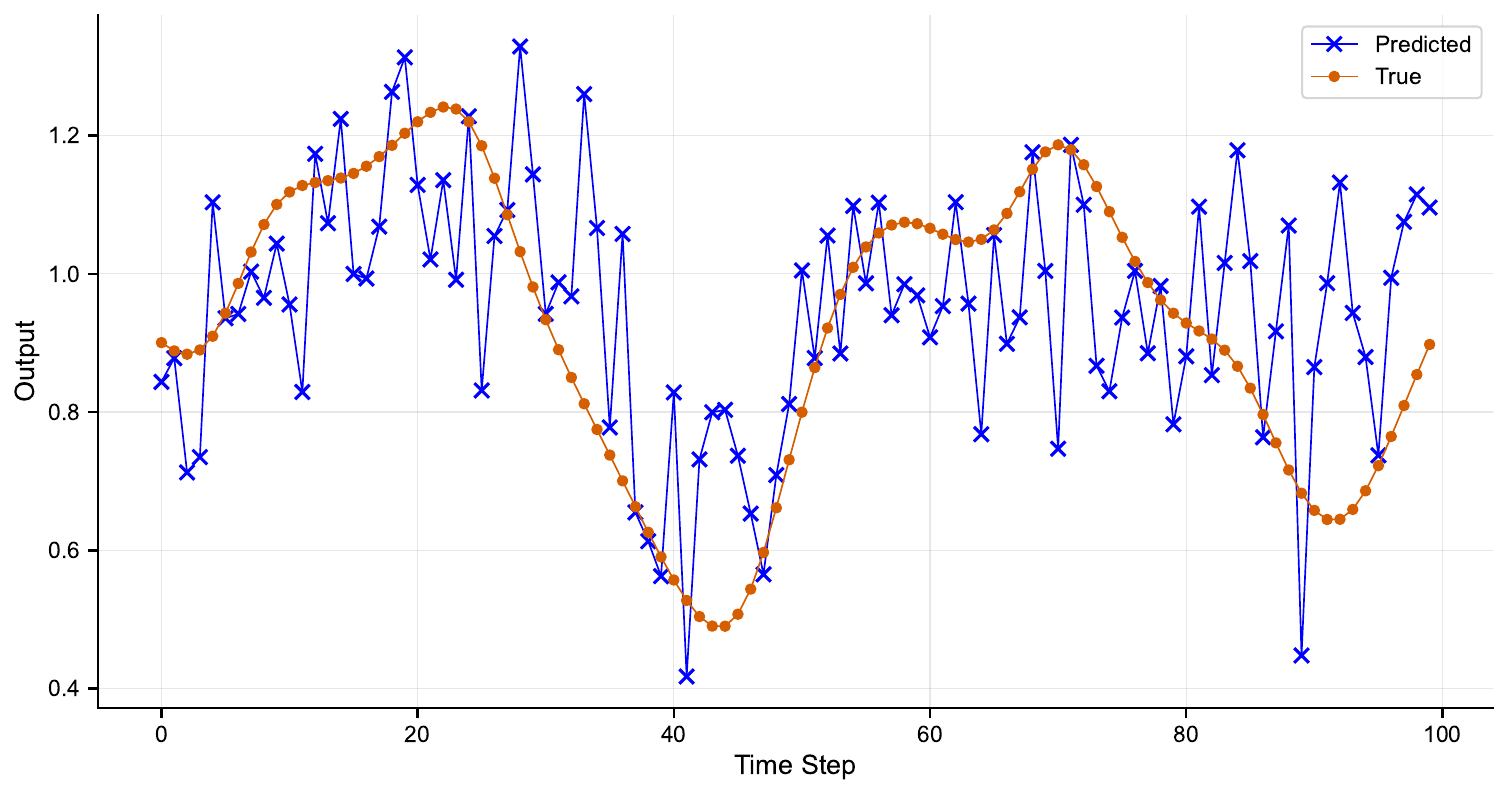}
    \caption{Predicted versus true Mackey-Glass signal for the quantum reservoir under noisy simulation at $\tau = 17$. The noisy prediction preserves the coarse oscillatory trend but exhibits strong local fluctuations and irregular spikes, indicating degradation of fine temporal structure under realistic device noise.}
    \label{fig:mse_vs_tau_noise}
\end{figure}

To assess the impact of realistic noise on the proposed quantum reservoir, we repeat the Mackey-Glass forecasting experiment using a noisy simulation based on a fake IBM\_Torino backend. The dataset consists of $5000$ samples with an 80/20 train-test split, window size $20$, prediction horizon $20$, and $\tau = 17$. As shown in Fig.~\ref{fig:mse_vs_tau_noise}, the QRC still captures the overall oscillatory structure of the signal. However, the predictions develop strong local fluctuations and irregular spikes, indicating that noise disrupts the fine temporal structure captured in the ideal simulation. This indicates that realistic device noise substantially disrupts the fine temporal structure learned by the reservoir, even when the coarse trend is still partially preserved. In particular, the noisy reservoir no longer matches the high-fidelity behavior observed in the ideal simulation, highlighting a clear gap between simulator performance and deployment under NISQ-like noise conditions. 

\begin{figure}[!h]
    \centering
    \includegraphics[width=\linewidth]{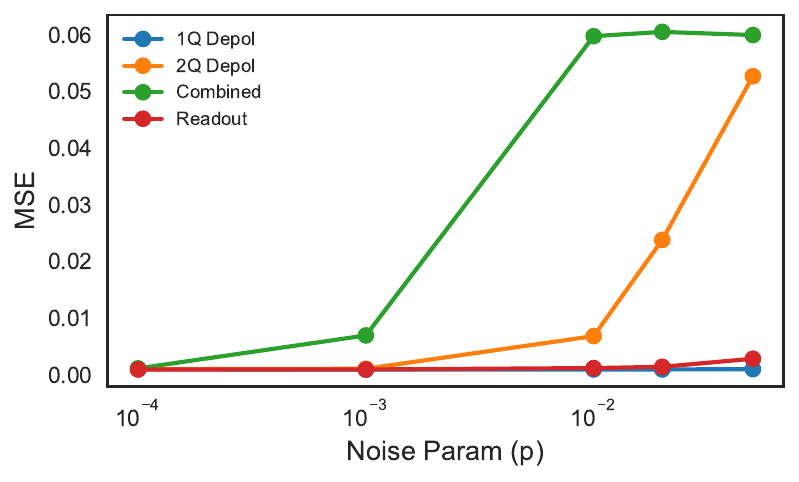}
    \caption{Performance (MSE score) of the quantum reservoir as a function of noise strength for different noise channels. 
    Single-qubit and readout noise have little effect, whereas two-qubit depolarizing noise leads to a sharp degradation. 
    The combined-noise case shows the strongest deterioration.}
    \label{fig:noise_param_combined}
\end{figure}

To further identify which noise mechanisms are most detrimental, we performed a targeted sensitivity analysis by varying individual noise channels separately. Figure~\ref{fig:noise_param_combined} summarizes the results in terms of mean squared error (MSE). We found that the reservoir is comparatively robust to single-qubit, readout, and relaxation noise over the tested range, whereas two-qubit gate noise leads to the strongest degradation in MSE value. When multiple noise sources are combined, the deterioration becomes even more severe, indicating that the instability observed in Fig.~\ref{fig:mse_vs_tau_noise} is driven primarily by entangling operations rather than by all noise channels equally.

\begin{figure}[!t]
    \centering
    \includegraphics[width=\linewidth]{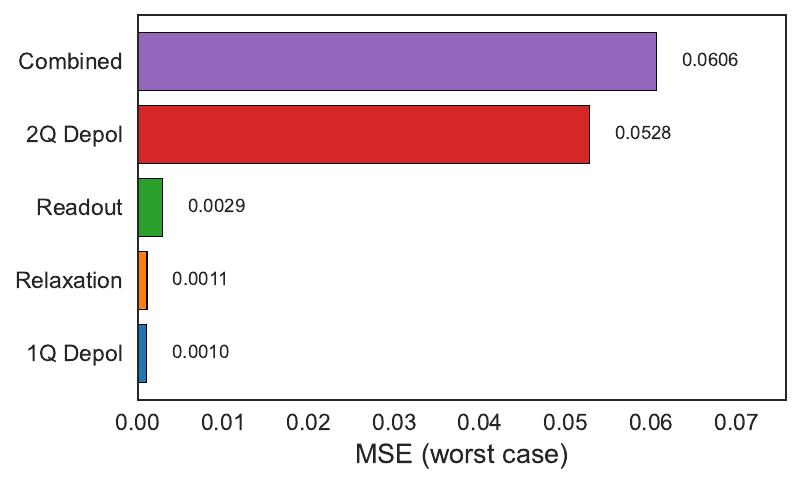}
    \caption{Worst-case MSE for each noise type at the maximum tested noise parameter. Two-qubit and combined noise dominate the degradation.}
    \label{fig:noise_ranking}
\end{figure}

For a compact comparison, Fig.~\ref{fig:noise_ranking} shows the worst-case MSE for each noise type. Two-qubit depolarizing noise and the combined-noise setting clearly dominate, while single-qubit, readout, and relaxation noise remain comparatively small.

This behavior aligns with the role of entanglement in the reservoir. The entangling gates that build the feature space also spread errors across the system, so noise introduced at these steps does not stay local but propagates through the dynamics. As a result, the reservoir loses both memory of past inputs and the structure needed for accurate prediction. This is consistent with the trends observed in Fig.~\ref{fig:entropy_memory_rmse}, where performance is tied to a balanced regime of entanglement and memory. In contrast, single-qubit and readout errors act more locally and therefore have a much weaker effect. Overall, these results point to entangling gate fidelity as the main constraint for running this model on near-term hardware.

\section{Discussion and Conclusion \label{conc_disc}}

Across the Mackey-Glass forecasting experiments, the proposed model achieves strong predictive performance and compares favorably with standard classical baselines. Notably, this performance is obtained using a fixed quantum circuit with no trainable internal parameters, indicating that the combination of structured feature-map encoding and a classical feedback loop is sufficient to generate expressive and stable reservoir dynamics. The results remain consistent across different delay parameters and prediction horizons, suggesting that the architecture generalizes well to varying temporal dependencies.

The results also clarify the role of feedback in the model. The performance sweeps over \(\alpha\) and \(\lambda\) show that good forecasting does not arise from circuit structure alone. Instead, it requires a balance between retaining past information through feedback and ensuring that enough of the reservoir state is accessible for prediction. Too little feedback weakens temporal memory, while excessive feedback can suppress useful input information. Likewise, truncating the readout too aggressively removes relevant dynamical features. The best performance is obtained in an intermediate regime, which is consistent with the broader reservoir computing principle that useful dynamics emerge when the system is neither too rigid nor too unstable.

This interpretation is reinforced by the dynamical analysis. The memory-capacity results show that the reservoir retains information over a finite but useful temporal window, while the echo-state tests confirm that the dynamics remain input-driven and stable. Together, these results indicate that the proposed architecture is not simply fitting the data through a large feature space, but is operating as a genuine reservoir with fading memory. The entanglement study further sharpens this picture: improved prediction is associated with parameter regions that support meaningful memory capacity and moderate entanglement, whereas very low or very high entanglement tends to be less favorable. In other words, the useful operating regime is not one of maximal quantum correlation, but one in which information spreading and dynamical stability are balanced.

At the same time, the noise study makes clear that the current performance of the model is much better in the ideal simulation than under realistic device-level noise. Although the reservoir remains relatively robust to single-qubit, readout, and relaxation noise, two-qubit errors lead to a considerable drop in performance, and combined noise quickly degrades the predictions. As a result, the model’s predictive quality is closely tied to the fidelity of entangling gates. This does not invalidate the architecture, but it does place an important qualification on near-term deployment: compact circuit design alone is not sufficient unless the hardware can support the required two-qubit operations with adequate accuracy.

Taken together, these results show that recurrent quantum feature maps provide a viable and conceptually clean route to quantum reservoir computing. The proposed hybrid feedback mechanism connects quantum feature maps, reservoir dynamics, and temporal learning in a single framework, while the accompanying analyses help explain why and when the model works. The main limitation identified here is sensitivity to two-qubit noise, which points naturally to future work on shallower entangling layouts, noise-aware feature-map design, and error-mitigation strategies. It will also be important to evaluate the architecture on a broader range of real-world temporal datasets and to study whether the same balance between memory, entanglement, and stability persists beyond Mackey-Glass. More broadly, this work suggests that the most promising route for QRC may not be to maximize quantum complexity at all costs, but to engineer structured quantum dynamics that are expressive enough to be useful and simple enough to remain stable on realistic hardware.

\section{Acknowledgments \label{secVI}}
The authors would like to acknowledge the use of IBM Quantum services for this work and, in particular, the Qiskit package \cite{Qiskit, qiskit-textbook}. AZG and KH acknowledge that the NRC headquarters is located on the traditional unceded territory of the Algonquin Anishinaabe and Mohawk people. K.H. acknowledges funding from the NSERC Discovery Grant. C.S. and K.H. would like to acknowledge the NRC for its Applied Quantum Computing Challenge Program and NSERC for the Alliance grant QIMMIQ.

\bibliography{refer}% Produces the bibliography via BibTeX.

@techreport{jaeger2001echo,
  title={The "echo state" approach to analysing and training recurrent neural networks-with an erratum note},
  author={Jaeger, Herbert},
  year={2001},
  institution={Bonn, Germany: German National Research Center for Information Technology GMD Technical Report}
}

@article{negoro2018machine,
  title={Machine learning with controllable quantum dynamics of a nuclear spin ensemble in a solid},
  author={Negoro, Makoto and Mitarai, Kosuke and Fujii, Keisuke and Nakajima, Kohei and Kitagawa, Masahiro},
  journal={arXiv preprint arXiv:1806.10910},
  year={2018}
}

@article{lukosevicius2009reservoir,
  title={Reservoir computing approaches to recurrent neural network training},
  author={Luko{\v{s}}evi{\v{c}}ius, Mantas and Jaeger, Herbert},
  journal={Computer Science Review},
  volume={3},
  number={3},
  pages={127--149},
  year={2009},
  publisher={Elsevier}
}

@article{jaeger2004harnessing,
  title={Harnessing nonlinearity: Predicting chaotic systems and saving energy in wireless communication},
  author={Jaeger, Herbert and Haas, Harald},
  journal={Science},
  volume={304},
  number={5667},
  pages={78--80},
  year={2004},
  publisher={American Association for the Advancement of Science},
  doi={10.1126/science.1091277},
  url={https://www.science.org/doi/10.1126/science.1091277}
}

@inproceedings{schrauwen2007overview,
  title={An overview of reservoir computing: theory, applications and implementations},
  author={Schrauwen, Benjamin and Verstraeten, David and Van Campenhout, Jan},
  booktitle={Proceedings of the 15th European Symposium on Artificial Neural Networks},
  pages={471--482},
  year={2007},
  url={https://www.elen.ucl.ac.be/Proceedings/esann/esannpdf/es2007-87.pdf}
}

@article{fujii2017harnessing,
  title={Harnessing disordered-ensemble quantum dynamics for machine learning},
  author={Fujii, Keisuke and Nakajima, Kohei},
  journal={Physical Review Applied},
  volume={8},
  number={2},
  pages={024030},
  year={2017},
  doi={10.1103/PhysRevApplied.8.024030},
  url={https://journals.aps.org/prapplied/abstract/10.1103/PhysRevApplied.8.024030}
}

@article{mujal2023time,
  title={Time-series quantum reservoir computing with weak and projective measurements},
  author={Mujal, Pere and Guerrero, Jorge and Garcia-Beni, David and Calvo, Gabriel F and Alarcon, Adria},
  journal={npj Quantum Information},
  volume={9},
  number={1},
  pages={16},
  year={2023},
  doi={10.1038/s41534-023-00682-z},
  url={https://www.nature.com/articles/s41534-023-00682-z}
}

@article{kobayashi2024feedback,
  title={Feedback-driven quantum reservoir computing for time-series analysis},
  author={Kobayashi, Keisuke and Fujii, Keisuke and Yamamoto, Yoshihisa},
  journal={PRX Quantum},
  volume={5},
  number={4},
  pages={040325},
  year={2024},
  doi={10.1103/PRXQuantum.5.040325},
  url={https://journals.aps.org/prxquantum/abstract/10.1103/PRXQuantum.5.040325}
}

@article{murauer2025feedback,
  title={Feedback Connections in Quantum Reservoir Computing with Mid-Circuit Measurements},
  author={Murauer, Thomas and Matsuura, Ayumu and Yamamoto, Yoshihisa},
  journal={arXiv preprint},
  year={2025},
  eprint={2503.22380},
  archivePrefix={arXiv},
  primaryClass={quant-ph},
  url={https://arxiv.org/abs/2503.22380}
}

@article{goto2021universal,
  title={Universal approximation property of quantum machine learning models in quantum-enhanced feature spaces},
  author={Goto, Hayata and Tran, Minh C. and Nakajima, Kohei},
  journal={Physical Review Letters},
  volume={127},
  number={9},
  pages={090506},
  year={2021},
  doi={10.1103/PhysRevLett.127.090506},
  url={https://doi.org/10.1103/PhysRevLett.127.090506}
}

@article{matsumoto2025iterative,
  title={Iterative quantum feature maps},
  author={Matsumoto, Keisuke and Tran, Minh C.},
  journal={arXiv preprint},
  year={2025},
  eprint={2506.19461},
  archivePrefix={arXiv},
  primaryClass={quant-ph},
  url={https://arxiv.org/abs/2506.19461}
}

@article{monomi2025weak,
  title={Feedback-enhanced quantum reservoir computing with weak measurements},
  author={Monomi, R. and Yamamoto, Y. and Fujii, K.},
  journal={arXiv preprint},
  year={2025},
  eprint={2503.17939},
  archivePrefix={arXiv},
  primaryClass={quant-ph},
  url={https://arxiv.org/abs/2503.17939}
}

@article{tanaka2019physical,
	author = {Tanaka, Gouhei and Yamane, Toshiyuki and H{\ifmmode\acute{e}\else\'{e}\fi}roux, Jean Benoit and Nakane, Ryosho and Kanazawa, Naoki and Takeda, Seiji and Numata, Hidetoshi and Nakano, Daiju and Hirose, Akira},
	title = {{Recent advances in physical reservoir computing: A review}},
	journal = {Neural Networks},
	volume = {115},
	pages = {100--123},
	year = {2019},
	month = jul,
	issn = {0893-6080},
	doi = {10.1016/j.neunet.2019.03.005},
        url= {https://www.sciencedirect.com/science/article/pii/S0893608019300784}
}

@article{maass2002real,
  title={Real-time computing without stable states: A new framework for neural computation based on perturbations},
  author={Maass, Wolfgang and Natschl{\"a}ger, Thomas and Markram, Henry},
  journal={Neural computation},
  volume={14},
  number={11},
  pages={2531--2560},
  year={2002},
  publisher={MIT Press}
}

@article{yildiz2012re,
  title={Re-visiting the echo state property},
  author={Yildiz, Izzet B and Jaeger, Herbert and Kiebel, Stefan J},
  journal={Neural networks},
  volume={35},
  pages={1--9},
  year={2012},
  publisher={Elsevier}
}

@article{goncalves2020reservoir,
  title={Reservoir computing universality with stochastic inputs},
  author={Gon{\c{c}}alves, Lyudmila and Marques, Nengli and Marques, Rui and Rego, Rodrigo},
  journal={IEEE Transactions on Neural Networks and Learning Systems},
  volume={31},
  number={11},
  pages={4749--4759},
  year={2020},
  publisher={IEEE}
}

@book{boyd1985fading,
  title={Fading memory and the problem of approximating nonlinear operators with Volterra series},
  author={Boyd, Stephen and Chua, Leon O},
  year={1985},
  publisher={IEEE Transactions on circuits and systems}
}

@techreport{jaeger2001short,
  title={Short term memory in echo state networks},
  author={Jaeger, Herbert},
  year={2002},
  institution={GMD-Forschungszentrum Informationstechnik}
}

@article{dambre2012information,
  title={Information processing capacity of dynamical systems},
  author={Dambre, Joni and Verstraeten, David and Schrauwen, Benjamin and Massar, Serge},
  journal={Scientific reports},
  volume={2},
  pages={514},
  year={2012},
  publisher={Nature Publishing Group}
}

@misc{Qiskit,
    author = {{Qiskit contributors}},
    title = {Qiskit: An Open-source Framework for Quantum Computing},
    year = {2023},
    doi = {10.5281/zenodo.2573505}
}

@book{qiskit-textbook,
  title = {Learn Quantum Computation Using Qiskit},
  author = {Abraham Asfaw and Luciano Bello and Yael Ben-Haim and Sergey Bravyi and Lauren Capelluto and Almudena Carrera Vazquez and Jack Ceroni and Richard Chen and Albert Frisch and Jay Gambetta and Shelly Garion and Leron Gil and Salvador De La Puente Gonzalez and Francis Harkins and Takashi Imamichi and David McKay and Antonio Mezzacapo and Zlatko Minev and Ramis Movassagh and Giacomo Nannicini and Paul Nation and Anna Phan and Marco Pistoia and Arthur Rattew and Joachim Schaefer and Javad Shabani and John Smolin and Kristan Temme and Madeleine Tod and Stephen Wood},
  year = {2020},
  url = {https://qiskit.org/textbook/preface.html}}

@article{nokkala2021gaussian,
  title={Gaussian states of continuous-variable quantum systems provide universal and versatile reservoir computing},
  author={Nokkala, Johannes et al.},
  journal={Communications Physics},
  volume={4},
  number={1},
  pages={53},
  year={2021}
}

@article{chen2020temporal,
  title={Temporal information processing on noisy quantum computers via reservoir computing},
  author={Chen, Jhen-Dong and Nurdin, Hendra I. and Yamamoto, Naoki},
  journal={Physical Review Applied},
  volume={14},
  number={2},
  pages={024065},
  year={2020}
}

@article{yasuda2023quantum,
  title={Quantum reservoir computing with repeated measurements on superconducting devices},
  author={Yasuda, Hiroshi and Yamamoto, Naoki},
  journal={arXiv preprint arXiv:2310.06706},
  year={2023}
}

@article{govia2021quantum,
  title={Quantum reservoir computing with a single nonlinear oscillator},
  author={Govia, Luke CG and others},
  journal={Physical Review Research},
  volume={3},
  number={1},
  pages={013077},
  year={2021}
}

@article{sannia2024dissipation,
  title={Dissipation as a resource for Quantum Reservoir Computing},
  author={Sannia, Antonio et al.},
  journal={Quantum},
  volume={8},
  pages={1291},
  year={2024}
}

@misc{quera2023scaling,
      title={Large-scale quantum reservoir learning with an analog quantum computer}, 
      author={Milan Kornjača and Hong-Ye Hu and Chen Zhao and Jonathan Wurtz and Phillip Weinberg and Majd Hamdan and Andrii Zhdanov and Sergio H. Cantu and Hengyun Zhou and Rodrigo Araiza Bravo and Kevin Bagnall and James I. Basham and Joseph Campo and Adam Choukri and Robert DeAngelo and Paige Frederick and David Haines and Julian Hammett and Ning Hsu and Ming-Guang Hu and Florian Huber and Paul Niklas Jepsen and Ningyuan Jia and Thomas Karolyshyn and Minho Kwon and John Long and Jonathan Lopatin and Alexander Lukin and Tommaso Macrì and Ognjen Marković and Luis A. Martínez-Martínez and Xianmei Meng and Evgeny Ostroumov and David Paquette and John Robinson and Pedro Sales Rodriguez and Anshuman Singh and Nandan Sinha and Henry Thoreen and Noel Wan and Daniel Waxman-Lenz and Tak Wong and Kai-Hsin Wu and Pedro L. S. Lopes and Yuval Boger and Nathan Gemelke and Takuya Kitagawa and Alexander Keesling and Xun Gao and Alexei Bylinskii and Susanne F. Yelin and Fangli Liu and Sheng-Tao Wang},
      year={2024},
      eprint={2407.02553},
      archivePrefix={arXiv},
      primaryClass={quant-ph},
      url={https://arxiv.org/abs/2407.02553}, 
}

@article{domingo2023noise,
  title={Taking advantage of noise in quantum reservoir computing},
  author={Domingo, Matteo et al.},
  journal={Scientific Reports},
  volume={13},
  number={1},
  pages={11591},
  year={2023}
}

@article{cindrak2024memory,
  title={Enhancing the performance of quantum reservoir computing and solving the time-complexity problem by artificial memory restriction},
  author={Čindrak, Luka et al.},
  journal={Physical Review Research},
  volume={6},
  number={1},
  pages={013076},
  year={2024}
}

@article{lukovsevivcius2009reservoir,
  title={Reservoir computing approaches to recurrent neural network training},
  author={Luko{\v{s}}evi{\v{c}}ius, Mantas and Jaeger, Herbert},
  journal={Computer Science Review},
  volume={3},
  number={3},
  pages={127--149},
  year={2009},
  publisher={Elsevier}
}

@misc{singh2025,
      title={A Resource Efficient Quantum Kernel}, 
      author={Utkarsh Singh and Jean-Frédéric Laprade and Aaron Z. Goldberg and Khabat Heshami},
      year={2025},
      eprint={2507.03689},
      archivePrefix={arXiv},
      primaryClass={quant-ph},
      url={https://arxiv.org/abs/2507.03689}, 
}

@article{havlivcek2019supervised,
	author = {Havl{\ifmmode\acute{\imath}\else\'{\i}\fi}{\ifmmode\check{c}\else\v{c}\fi}ek, Vojt{\ifmmode\check{e}\else\v{e}\fi}ch and C{\ifmmode\acute{o}\else\'{o}\fi}rcoles, Antonio D. and Temme, Kristan and Harrow, Aram W. and Kandala, Abhinav and Chow, Jerry M. and Gambetta, Jay M.},
	title = {{Supervised learning with quantum-enhanced feature spaces}},
	journal = {Nature},
	volume = {567},
	number = {7747},
	pages = {209--212},
	year = {2019},
	month = mar,
	issn = {1476-4687},
	publisher = {Nature Publishing Group},
	doi = {10.1038/s41586-019-0980-2}
}

@article{verstraeten2007experimental,
title = {An experimental unification of reservoir computing methods},
journal = {Neural Networks},
volume = {20},
number = {3},
pages = {391-403},
year = {2007},
note = {Echo State Networks and Liquid State Machines},
issn = {0893-6080},
doi = {https://doi.org/10.1016/j.neunet.2007.04.003},
url = {https://www.sciencedirect.com/science/article/pii/S089360800700038X},
author = {D. Verstraeten and B. Schrauwen and M. D’Haene and D. Stroobandt},
}

@article{mackey1977oscillation,
  title={Oscillation and chaos in physiological control systems},
  author={Mackey, Michael C. and Glass, Leon},
  journal={Science},
  volume={197},
  number={4300},
  pages={287--289},
  year={1977},
  publisher={American Association for the Advancement of Science},
  doi={10.1126/science.267326},
  url={https://doi.org/10.1126/science.267326}
}

@article{PRXQuantum_feedback,
  title = {Feedback-Driven Quantum Reservoir Computing for Time-Series Analysis},
  author = {Kobayashi, Kaito and Fujii, Keisuke and Yamamoto, Naoki},
  journal = {PRX Quantum},
  volume = {5},
  issue = {4},
  pages = {040325},
  numpages = {16},
  year = {2024},
  month = {Nov},
  publisher = {American Physical Society},
  doi = {10.1103/PRXQuantum.5.040325},
  url = {https://link.aps.org/doi/10.1103/PRXQuantum.5.040325}
}

@misc{gonon2026feedbackdrivenrecurrentquantumneural,
      title={Feedback-driven recurrent quantum neural network universality}, 
      author={Lukas Gonon and Rodrigo Martínez-Peña and Juan-Pablo Ortega},
      year={2026},
      eprint={2506.16332},
      archivePrefix={arXiv},
      primaryClass={quant-ph},
      url={https://arxiv.org/abs/2506.16332}, 
}

@article{grigoryeva2018echo,
  author = {L. Grigoryeva and J.-P. Ortega},
  title = {Echo state networks are universal},
  journal = {Neural Networks},
  volume = {108},
  pages = {495--508},
  year = {2018},
  doi = {10.1016/j.neunet.2018.08.025},
  url = {https://doi.org/10.1016/j.neunet.2018.08.025}
}

\newpage
\clearpage

% \onecolumngrid
\begin{appendix}

\end{appendix}
\subsection{Effect of Feedback Strength and Readout Truncation \label{fedback_trunc}}
We next examine how two core parameters—feedback strength \(\alpha\) and readout limit \(\lambda\)—influence forecasting performance. As described previously, \(\alpha\) controls the contribution of the feedback vector \(\tilde{\mathbf{z}} = \alpha \cdot \mathbf{z}\), where \(\mathbf{z}\) is the output of the previous reservoir iteration. Setting \(\alpha = 0\) removes recurrence entirely, while \(\alpha = 1\) reuses the full previous output. The parameter \(\lambda \in (0,1]\) determines the fraction of output amplitudes used in the readout vector after each quantum circuit execution. For instance, \(\lambda = 0.6\) means only the first 60\% of states—ranked lexicographically—are retained as features for regression.

\begin{figure}[H]
    \centering
    % Top Figure
    \includegraphics[width=\linewidth]{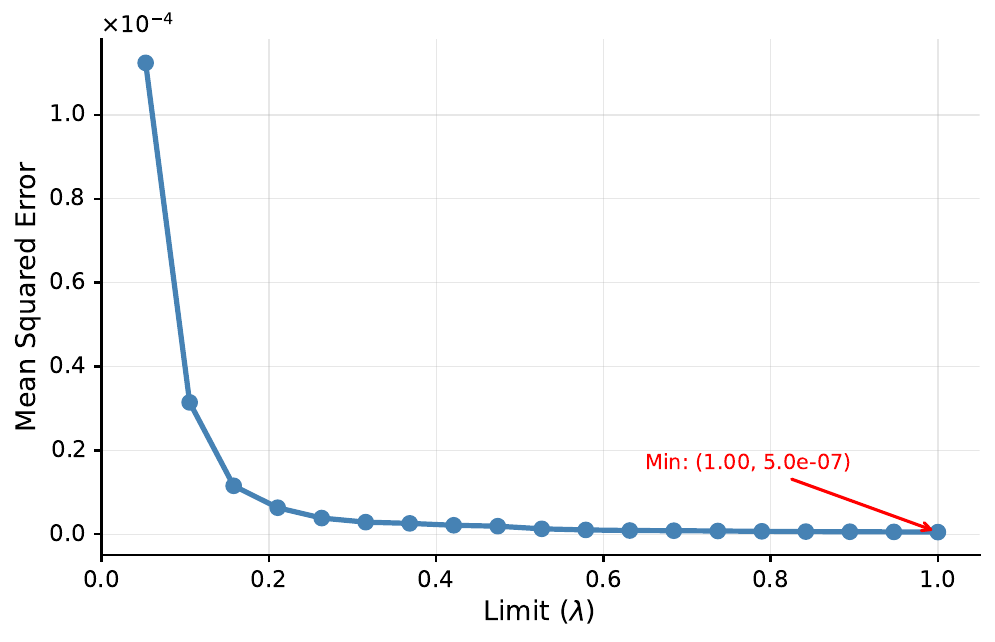}
    \\[1ex] % Adds a vertical gap between images

    % Bottom Figure
    \includegraphics[width=\linewidth]{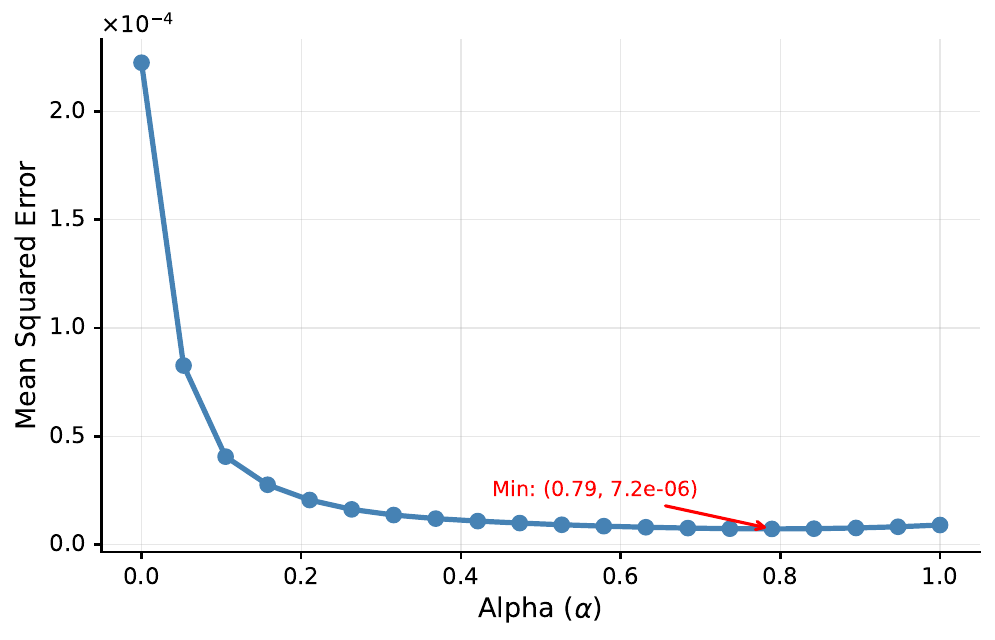}
    
    \caption{Variation of test MSE with readout limit \(\lambda\) (top) and feedback strength \(\alpha\) (bottom). The best performance is achieved at \((\alpha, \lambda) = (0.79, 1.0)\), demonstrating the importance of both recurrence and output dimensionality in quantum reservoir dynamics.}
    \label{fig:qrc_lambda_alpha_vs_mse}
\end{figure}

Figure~\ref{fig:qrc_lambda_alpha_vs_mse} summarizes the effect of both parameters on mean squared error (MSE), evaluated on the Mackey--Glass dataset with \(\tau = 17\), window size = 20, and prediction horizon = 20. The model achieves its best performance at \((\alpha, \lambda) = (0.79, 1.0)\), indicating that strong—but not maximal—feedback and full readout provide optimal dynamics for this task. Lower values of \(\alpha\) degrade memory, while overly large \(\alpha\) may cause feedback dominance and input suppression. Similarly, reducing \(\lambda\) discards useful dynamical information, limiting the reservoir’s expressive capacity. 

These results highlight that performance in feedback-driven quantum reservoirs arises not only from circuit design, but also from a careful balance between recurrence and observability.

\subsection{Memory Dynamics and Echo State Property \label{mc_esp}}

\begin{figure}[H]
    \centering
    \begin{subfigure}[t]{0.48\textwidth}
        \centering
        \includegraphics[width=\linewidth]{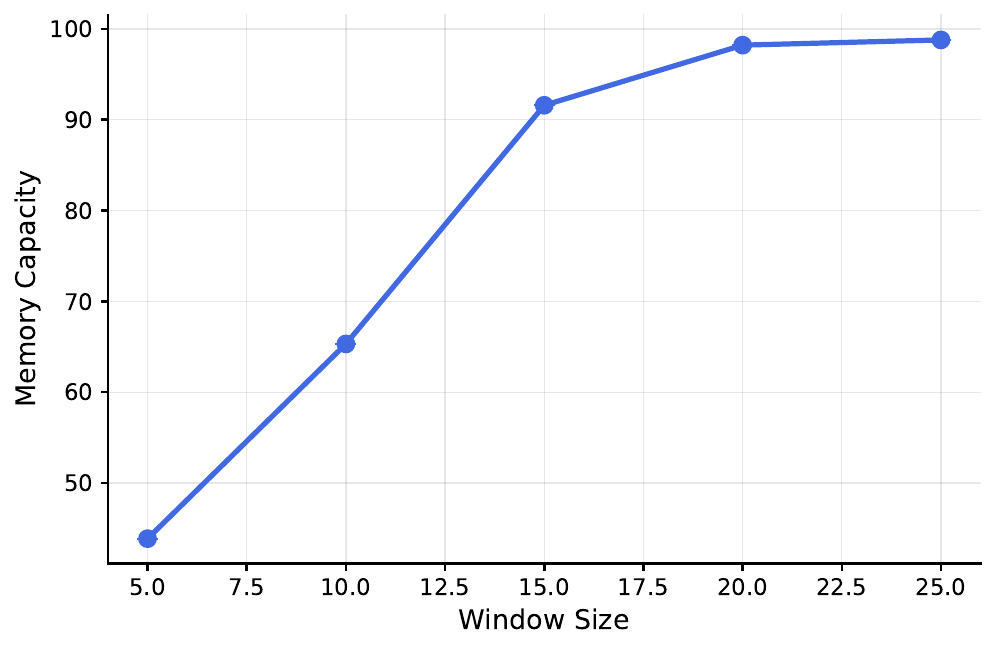}
        \caption{Memory capacity vs. input window size.}
    \end{subfigure}
    \hfill
    \begin{subfigure}[t]{0.48\textwidth}
        \centering
        \includegraphics[width=\linewidth]{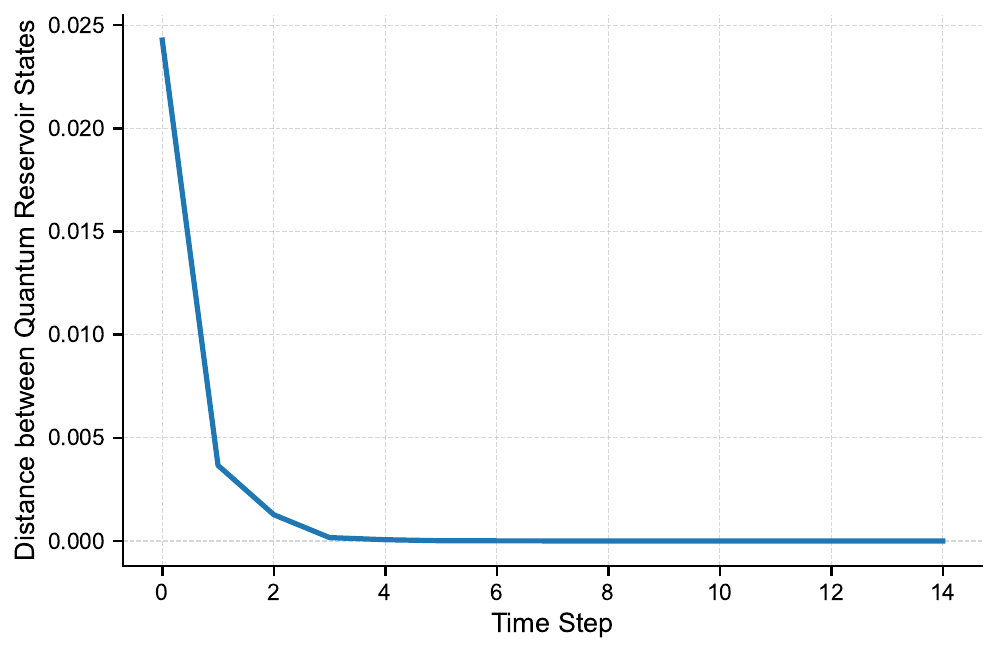}
        \caption{State distance over time for ESP verification.}
    \end{subfigure}
    \caption{Dynamical analysis of the quantum reservoir. (a) Memory capacity increases with window size and saturates around 20, indicating a limit in temporal retention. (b) Convergence of reservoir trajectories confirms that the quantum system satisfies the echo state property.}
    \label{fig:memory_esp}
\end{figure}

We assess two fundamental dynamical properties of our quantum reservoir system: its linear memory capacity and its stability under the echo state property (ESP). The memory capacity quantifies how well past inputs can be reconstructed from the current reservoir state, reflecting the system’s ability to retain temporal information. As shown in Figure~\ref{fig:memory_esp}, the memory capacity increases with window size and begins to saturate around a window length of 20, indicating that the reservoir effectively retains information over several time steps without being overwhelmed. Beyond this point, 

 additional inputs offer diminishing returns, consistent with capacity saturation observed in classical echo state networks~\cite{jaeger2001short, dambre2012information}.

To evaluate stability, we measure the state distance between two initially different reservoir trajectories driven by identical input sequences. The results confirm that, after a short transient, the state distance diminishes and converges toward zero, demonstrating that the reservoir dynamics are input-driven rather than history-dependent. This convergence behavior verifies that our quantum reservoir satisfies the ESP, a critical condition for consistent temporal processing and fading memory. Together, these findings validate the use of our feedback-based quantum reservoir as a stable and memory-efficient temporal model.

\subsection{Performance of QRC in presence of relaxation noise:}
\begin{figure}[!h]
    \centering
    \includegraphics[width=\linewidth]{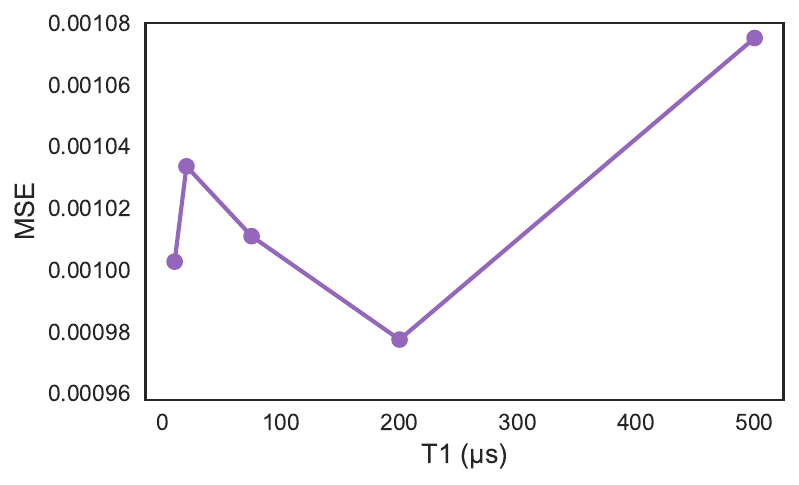}
    \caption{MSE as a function of relaxation time $T_1$. The weak dependence on $T_1$ indicates that relaxation noise has a limited impact on performance in the explored regime.}
    \label{fig:rmse_t1_relaxation}
\end{figure}

We also assessed the performance in the presence of relaxation noise. As shown in Fig.~\ref{fig:rmse_t1_relaxation}, the MSE changes very little over a wide range of $T_1$ values, which suggests relaxation alone is not the main factor limiting performance in this setting.

\subsection{Full-state feedback:}

\begin{figure}[htb]
    \centering
    \includegraphics[width=\linewidth]{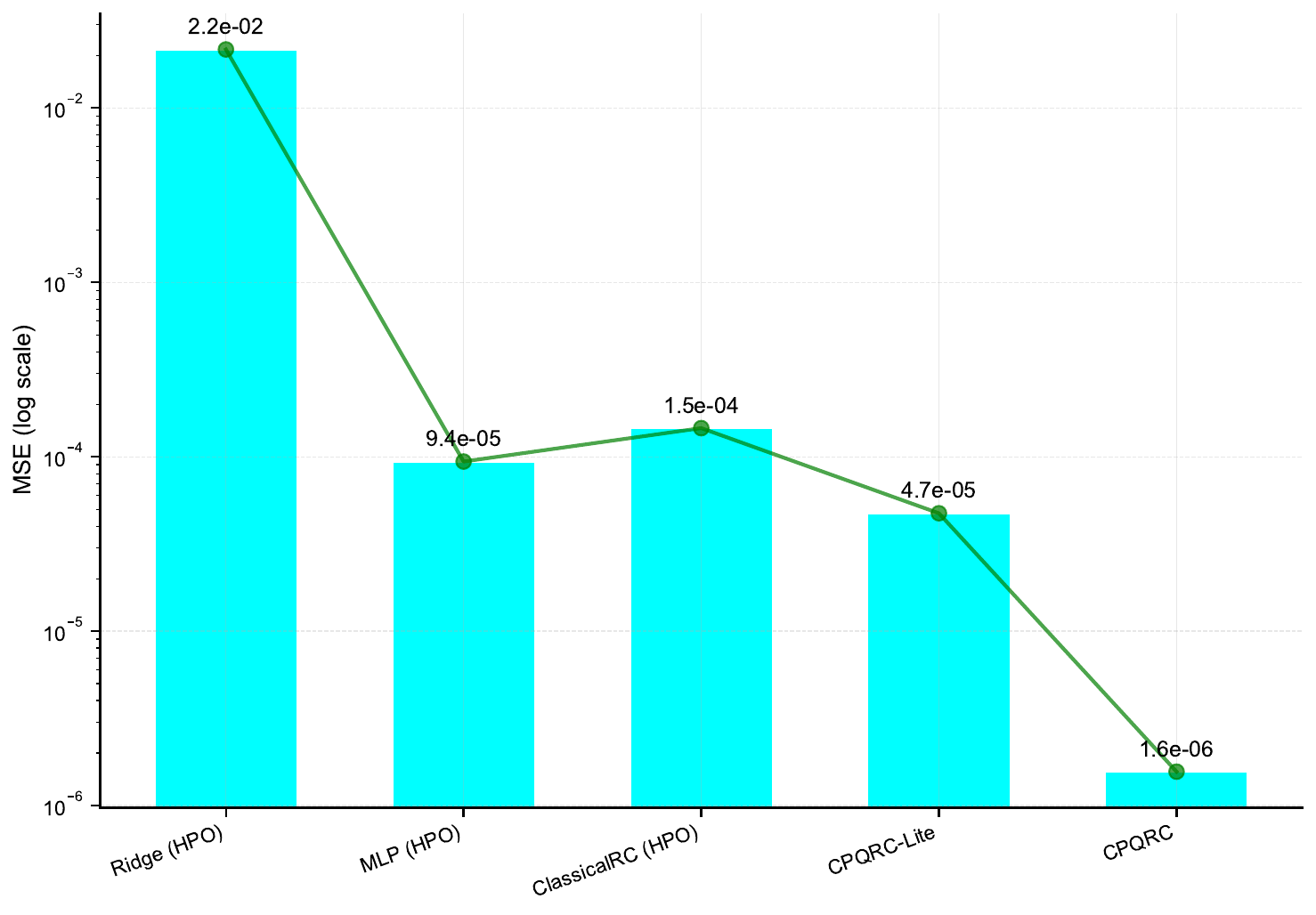}
    \caption{Model comparison on the Mackey--Glass dataset at \(\tau = 17\), window size = 20, prediction horizon = 20. CPQRC-lite achieves lower error than classical reservoir computing and performs comparably to MLP, while the original CPQRC model attains the lowest MSE.}
    \label{fig:cpqrc_lite_feedback_comparison}
\end{figure}

We evaluate the full-state feedback variant (CPQRC-lite) on the Mackey--Glass time-series prediction task using 5000 samples, a window size of 20, prediction horizon of 20, and delay parameter $\tau = 17$. The results are summarized in Fig.~\ref{fig:cpqrc_lite_feedback_comparison} and compared with standard baselines, including Ridge regression, MLP, classical reservoir computing, and the original CPQRC model. Here, CPQRC represents the QRC framework proposed in the paper with the CPMap circuit.

\renewcommand{\thealgocf}{} 
\begin{algorithm}[htbp]
\caption{Feedback-Driven Quantum Reservoir Computing without single-qubit expectations}
\label{alg:qrc_full_feedback}
\DontPrintSemicolon

\KwIn{Time-series input $\{x_t\}_{t=1}^{T}$, feature map $U(\cdot)$, window size $\tau$, feedback strength $\alpha$, output fraction $\lambda$, prediction horizon $H$}

\KwOut{Predicted outputs $\{\hat{y}_t\}_{t=\tau+1}^{T-H}$}

Initialize feedback vector $\tilde{\mathbf{z}}_0 \leftarrow \mathbf{0} \in \mathbb{R}^{n_q}$\;

\For{$t = \tau+1$ \KwTo $T - H$}{

    Create input window: $\mathbf{x}_t = [x_{t-\tau}, \dots, x_t]$\;

    Encode input with feature map: apply $U(\mathbf{x}_t)$ to left half of the circuit\;

    Encode feedback from previous step: apply $U^\dagger(\Pi(\tilde{\mathbf{z}}_{t-1}))$ to right half\;

    Execute full quantum circuit and measure output distribution $\mathbf{p}_t$\;

    Extract first $n = \mathrm{dim}(\mathbf{x}_t)$ components from $\mathbf{p}_t$:
        $\mathbf{z}_t = (p_t(s_1), \dots, p_t(s_n))$;

    Update feedback: $\tilde{\mathbf{z}}_t = \alpha \cdot \mathbf{z}_t$\;

    Construct regression features:
    $\mathbf{r}_t = \bigl(p_t(s_1), \dots, p_t(s_{\lfloor \lambda 2^{n_q} \rfloor})\bigr)$\;

    Store $(\mathbf{r}_t, y_t)$ where $y_t = x_{t+H}$\;
}

Train regression model: $\hat{y}_t = \mathbf{w}_{\mathrm{out}}^\top \mathbf{r}_t + b$\;

\Return predicted sequence $\{\hat{y}_t\}_{t=\tau+1}^{T-H}$

\end{algorithm}

CPQRC-lite achieves a mean squared error of $4.7 \times 10^{-5}$, which is lower than classical reservoir computing ($1.5 \times 10^{-4}$) and comparable to the MLP baseline ($9.4 \times 10^{-5}$). The original CPQRC model achieves the lowest error of $1.6 \times 10^{-6}$. These results show that the full-state feedback approach maintains competitive performance while simplifying the feedback construction by directly using the measurement distribution, without requiring explicit computation of expectation values.

\end{document}